\documentclass[11pt]{article}
\usepackage{fullpage}
\usepackage{amsmath}
\usepackage{amssymb}
\usepackage[dvips]{epsfig}
\usepackage{epstopdf}
\usepackage{color}
\usepackage{cite}
\usepackage{upgreek}

\def\##1{\underline #1}
\def\=#1{\underline{\underline #1}}

\def\eps{\epsilon}
\def\epso{\epsilon_0}
\def\muo{\mu_0}
\def\ko{k_0}
\def\kosq{k_0^2}
\def\lambdao{\lambda_0}
\def\etao{\eta_0}
\def\.{\mbox{ \tiny{$^\bullet$} }}

\def\epsa{\epsilon_{a}}
\def\epsb{\epsilon_{b}}
\def\epsc{\epsilon_{c}}
\def\epsdt{\tilde{\epsilon}_{d}}
\def\epsrefo{\=\eps_{\,ref}^o}
\def\epss{\epsilon_{s}}
\def\piom{\frac{\pi}{\Omega}}

\def\ux{\#{u}_x}
\def\uy{\#{u}_y}
\def\uz{\#{u}_z}
\def\up{\#{u}_+}
\def\um{\#{u}_-}

\def\aal{a_L}
\def\aar{a_R}
\def\bbl{r_L}
\def\bbr{r_R}

\def\rrr{r_{RR}}

\def\le{\left(}
\def\ri{\right)}
\def\les{\left[}
\def\ris{\right]}
\def\lec{\left\{}
\def\ric{\right\}}

\def\c#1{\cite{#1}}

\def\r#1{(\ref{#1})}

\begin{document}
%\noindent Submitted for publication in {\it Microwave \& Optical Technology Letters}
%\vskip 0.4cm

\begin{center}
{\large {\bf On Percolation and Circular Bragg Phenomenon in Metallic, 
Helicoidally Periodic, Sculptured Thin Films}}
\vskip 0.4cm

\noindent  {\bf Akhlesh Lakhtakia\/}\footnote{Tel: +1 814 863 4319; Fax: +1 814 863 7967; e-mail: AXL4@psu.edu\\} 
\vskip 0.2cm
\noindent {\em CATMAS~--~Computational \& Theoretical Materials Sciences Group\\
Department of Engineering Science \& Mechanics\\
Pennsylvania State University, University Park, PA 16802--1401, USA}

\end{center}

\vskip 1.0cm

\noindent {\bf Abstract:} The concept of local homogenization is invoked 
in the visible and the sub-visible frequency regimes to
estimate the permittivity dyadics of a thin-film helicoidal bianisotropic
mediums (TFHBMs)~---~a canonical
class of sculptured thin films with helicoidally
periodic microstructure~---~made
of a metal. Examination of the components of the permittivity dyadics,
as well as of the co- and the cross-polarized reflectances of axially excited, metallic
TFHBM halfspaces, reveals that three different absorbing-dielectric-to-metal
transitions will occur as the metallic volumetric fraction is increased,
but only two will be optically sensed using normally incident plane waves.
The first transition is show to virtually eliminate the circular Bragg phenomenon 
characteristically displayed by axially excited, dielectric TFHBMs.

\vskip 0.2cm


\section{Introduction}

Microstructurally, a sculptured thin film (STF) is best visualized as 
an assembly of parallel columns that all curve in the
same way in the thickness direction \c{LM97}.
Thus, a STF is locally anisotropic and unidirectionally nonhomogeneous,
and may be considered as a nonhomogeneous continuum in the visible
and the sub-visible frequency regimes.
The aim of this communication is to theoretically
delineate the effect of percolation
on the circular Bragg phenomenon (CBP) displayed by the so-called
thin-film helicoidal bianisotropic mediums (TFHBMs)~---~a canonical
class of STFs with helicoidally
periodic microstructure~---~made
of metals. 

Metal-dielectric particulate composite materials are known to demonstrate the
phenomenon of percolation \c{Sheng} in the following manner:
An insulator-to-conductor transition occurs abruptly
as $f$, the volumetric fraction of metallic particles, increases
from zero. Provided the metallic 
particles are spherical, the transition occurs at $f =f_{pt} \sim 0.33$,
where $f_{pt}$ is the percolation threshold. Different values
of $f_{pt}$ are found for nonspherical particles, because it is
a strong function of particulate geometry and orientational
statistics. This was exemplified by
Lakhtakia {\em et al.\/} \c{LMW97} for uniaxial composite mediums
formed by randomly dispersing  parallel,   aciculate (i.e., needle-like),
dielectric  particles
in some isotropic dielectric host medium.

One of the two Bruggeman models set up for the homogenization of
uniaxial composite mediums \c{LMW97} is also the basis
of a simple model to predict the effective constitutive properties
of  dielectric TFHBMs \c{SLM}, wherein
the continuous columns are modeled
as ensembles of needles.  It stands to reason  that percolation
must be exhibited by metallic TFHBMs. Although a few metallic TFHBMs
have been fabricated \c{RB97, LUS99},  their optical responses
remain unexplored. The lack of experimental
data may be due to the structural irregularities in early specimen of
metallic TFHBMs \c{RB97}, the causes of and the remedies for irregularities
being topics of current research \c{LUS99}-\c{LM99}. These
laboratory fabrication 
efforts motivated the theoretical study reported here.

The plan of this paper is as follows: Section 2 contains the
theoretical preliminaries. The permittivity dyadic of a
metallic TFHBM is presented, and a simple homogenization model to estimate
the relative permittivity scalars is briefly recounted. In Section 3,
a boundary value problem is formulated for the reflection of
a normally incident plane wave that excites a metallic TFHBM 
halfspace along its axis of helicoidal periodicity. The effect
of percolation on the planewave response of the metallic TFHBM halfspace
is addressed in Section 3. Vectors are underlined, dyadics are double-underlined;
while an $\exp(-i\omega t)$ time-dependence is implicit,
with $\omega$ as the angular frequency.

\section{Theoretical Preliminaries}
 
\subsection{Permittivity dyadic}
Although all previous investigations on STFs considered
the reflection/transmission responses of films of
infinite lateral extent and finite width, determination of the response
of a metallic TFHBM halfspace is
more appropriate in the present context. This is because
excessive attenuation inside a TFHBM layer of finite width
leads to virtually zero fields on the layer's back face.
The exact solution of a relevant
two-point boundary value problem \c{VL98a} is then
severely compromised for even a moderately thick TFHBM layer.

Suppose the halfspace $z \geq 0$ is occupied
by a metallic TFHBM, while the halfspace $z \leq 0$ is vacuous. 
The linear dielectric properties of the TFHBM are delineated by the
nonhomogeneous  permittivity dyadic
\begin{equation}
\=\eps(\#r) = \epso\, \=S_z(z,h)\.\=S_y(\chi)\.\epsrefo
\.\=S_y^{-1}(\chi)\.
\=S_z^{-1}(z,h)\, ; \qquad z \geq 0\, ,
\label{epsbasic}
\end{equation}
where
\begin{equation}
\label{epsref}
\epsrefo= \epsa \,\uz\uz +\epsb\,\ux\ux+\epsc\,\uy\uy\, 
\end{equation}
is called  the {\em local\/} relative permittivity dyadic.
The relative
permittivity scalars $\eps_{a,b,c}$ are implicit functions
of frequency, and the presented analysis applies to dielectric
TFHBMs too.
Here and hereafter, $\epso = 8.854\times 10^{-12}$~F/m and
$\muo=4\pi\times 10^{-7}$~H/m are the permittivity and
the permeability of free space (i.e., vacuum), respectively; $\ko =  2\pi/\lambdao
=\omega\,\sqrt{\muo\epso}$  and $\lambdao$ are the wavenumber
and wavelength in free space, respectively; $\etao=\sqrt{\muo/\epso}$ is the intrinsic
impedance of free space; while $\ux$, $\uy$ and $\uz$
denote the unit vectors in a cartesian coordinate system. 

The helicoidal periodicity of the chosen TFHBM is captured by the
rotation dyadic
\begin{eqnarray}
\nonumber
\=S_z(z,h) &=&
\uz\uz + \le \ux\ux+\uy\uy\ri\,\cos\frac{\pi z}{\Omega} \\ 
& & +\,\, h\,\le \uy\ux-\ux\uy\ri\,\sin\frac{\pi z}{\Omega}\, .
\end{eqnarray}
The direction of the nonhomogeneity is parallel to the $z$ axis,
with $2\Omega$ as the structural period. The integer $h=1$ for a
structurally right-handed TFHBM; and $h=-1$ for structural left-handedness.
 The tilt dyadic
\begin{equation}
\=S_y(\chi) = \uy\uy + (\ux\ux + \uz\uz) \, \cos\chi + (\uz\ux-\ux\uz)\,
\sin\chi
\end{equation}
represents the {\it locally\/} aciculate microstructure of a TFHBM,
with $\chi > 0^\circ$ being the tilt angle \c{SLM,ML99}.

\subsection{Estimation of $\eps_{a,b,c}$}
Let the chosen TFHBM be fabricated of an isotropic material whose
relative permittivity in the bulk state is denoted by $\eps_s$
at the frequency of interest. The columns are supposed
to comprise identical long needles,
whose length is $u$ $(\gg 1)$ times greater than their cross-sectional radius.
Additionally, without loss of generality in
the present context,  the cross-section of the needles
is assumed to be circular, which implies that $\eps_c=\eps_a$
in \r{epsref}.
The needles are electrically small, the film is
porous, and the void regions are vacuous. The orientation of
the needles changes with $z$, but not with $x$ and $y$. The
TFHBM can thus be viewed  {\em locally\/}  as a two-phase composite material
with the so-called 3-1 connectivity \c{NSC}, as the vacuous phase
is connected in 3-D and the metallic phase in 1-D. Of course,
the 1-D connectivity of the metallic phase twists with $z$,
in accordance with the helix described by the unit vector
\begin{equation}
\#u_t(z)= \=S_z(z,h)\.\=S_y(\chi)\.\ux\, .
\end{equation}

The concept of local homogenization \c{SLM}
is invoked now to estimate the effective dielectric response
properties of the chosen TFHBM.
Application of the Bruggeman formalism leads
to the dyadic equation \c{LMW97,SLM}
\begin{eqnarray}
\nonumber
 \=0 &=& (1-f) \, \les \epsrefo -  \=I \ris \.
\lec \=I - \=W\. \les \=I -  {\epsrefo}^{-1} \ris \ric^{-1} 
\\
\label{brug}
&+&\,  f \, \les \epsrefo - \eps_s \=I \ris \.
\lec \=I - \=W\. \les \=I - \eps_s \, {\epsrefo}^{-1} \ris
\ric^{-1} \, ,
\end{eqnarray}
where the shape factor $u$ of the needles occurs in the
dyadic
\begin{equation}
\=W = \frac{1}{2}\, \les 1 + 4 u^{-2}\,\frac{\epsb}{\epsa}
\ris^{-1/2} \, \=I
+ \,\lec 1 - \frac{3}{2}\, \les 1 + 4 u^{-2}\,\frac{\epsb}{\epsa}
\ris^{-1/2} \ric \, \ux\ux \, ;
\end{equation}
the metallic volumetric fraction is denoted
by $f$, ($0\leq f \leq 1$); while $\=I$
and $\=0$ are the unit and the null dyadics, respectively.
Equation \r{brug} has to be solved numerically to estimate
the constituents $\epsa$ and $\epsb$ of $\epsrefo$.

\section{Boundary Value Problem}
Suppose an arbitrarily polarized plane wave is
normally incident on the chosen
TFHBM halfspace from the lower halfspace $z \leq 0$.
As a result of the axial excitation
of the upper halfspace, a plane wave is reflected into the lower halfspace.
The electric field phasor associated with the two plane
waves in the lower halfspace is stated as \c{VL98a}
\begin{eqnarray}
\nonumber
\#E(z) &=&  \le  \aal \, \up + \aar \, \um \ri \,
\exp\le i \ko z \ri\\
&+&\, \le  \bbl \, \um + \bbr \, \up \ri \,
\exp\le -i \ko z \ri \, ;  \qquad z \leq 0\, ,
\end{eqnarray}
and the corresponding magnetic field phasor
is then easily determined from the Faraday equation.
Here, the complex unit vectors $\#u_\pm = (\ux \pm i \uy)/\sqrt{2}$;
$\aal$ and $\aar$ are the known amplitudes
of the left-- and the right--circularly polarized (LCP \& RCP)
components
of the incident plane wave; and
$\bbl$ and $\bbr$ are the unknown amplitudes 
of the reflected planewave components. 
Our intention is to
determine the reflection coefficients
entering the 2$\times$2 matrix in the following relation:
\begin{equation}
\label{eq15}
\les \begin{array}{cccc} \bbl \\ \bbr  \end{array}\ris  =
\les \begin{array}{cccc} r_{LL} & r_{LR} \\ r_{RL} & r_{RR}\end{array}\ris \,
\les \begin{array}{cccc} \aal \\ \aar  \end{array}\ris \, .
\end{equation}
These coefficients are doubly subscripted:
those with both subscripts identical refer to co-polarized,
while those with two different subscripts denote
cross-polarized, reflection. 

The specification of fields induced in the TFHBM halfspace
requires some care. Four modes can propagate in the 
$\pm z$ direction; therefore \c{L99},
\begin{equation}
\label{efield0}
\begin{array}{l}
\#E(z) = 
{\sum\limits_{n=1}^{4}}{'}\, a_n\, e^{ig_nz} \lec
\ux\, \les e_{n1}\, \cos(\pi z/\Omega) - e_{n2}\, \sin(\pi z/\Omega) \ris \right.\\ [5pt]
\qquad\qquad + \left.
\uy\, \les e_{n1} \,\sin(\pi z/\Omega) + e_{n2} \,\cos(\pi z/\Omega) \ris +
\uz\,e_{n3}\ric \, ;  \qquad z \geq 0\, ,
\end{array}
\end{equation}
and
\begin{equation}
\label{hfield0}
\begin{array}{l}
\#H(z) = 
{\sum\limits_{n=1}^{4}}{'}\, a_n\, e^{ig_nz} \lec
\ux\, \les h_{n1}\, \cos(\pi z/\Omega) - h_{n2}\, \sin(\pi z/\Omega) \ris \right.\\ [5pt]
\qquad\qquad + \left.
\uy\, \les h_{n1} \,\sin(\pi z/\Omega) + h_{n2} \,\cos(\pi z/\Omega) \ris \ric \, ;  \qquad z \geq 0\, .
\end{array}
\end{equation}
The (un--normalized) cartesian 
components of the modal field phasors, given by \c{VL98a, L99}
\begin{equation}
\left. \begin{array}{l}
e_{n1} = \omega\muo \,\les g_n^2 - \kosq\epsc + (\piom)^2\ris\\ [8pt]
e_{n2} = 2i\omega\muo \,\piom\,g_n\\ [8pt]
e_{n3}= e_{n1}\, \frac{\epsa-\epsb}{\epsa\epsb}\,\epsdt\,\cos\chi\,\sin\chi\\[8pt]
h_{n1} = -i\,\piom\, \les g_n^2 + \kosq\epsc - (\piom)^2\ris\\ [8pt]
h_{n2} = g_n\, \les g_n^2 - \kosq\epsc - (\piom)^2\ris
\end{array}\ric\, ; \quad 1 \leq n \leq 4\, ,
\end{equation}
contain the four modal wavenumbers 
\begin{equation}
\label{g1}
\begin{array}{l}
g_1 = -g_3 = +\,2^{-1/2}\,\lec
\kosq\, (\epsc+\epsdt) + 2\,(\piom)^2 \right.\\[5pt]
\qquad\qquad + \left.\ko \les \kosq(\epsc-\epsdt)^2 + 8\,(\piom)^2
\,(\epsc+\epsdt)\ris^{1/2}
\ric^{1/2}\, ,
\end{array}
\end{equation}
\begin{equation}
\label{g2}
\begin{array}{l}
g_2 = -g_4 = +\,2^{-1/2}\,\lec
\kosq\, (\epsc+\epsdt) + 2\,(\piom)^2 \right.\\[5pt]
\qquad\qquad - \left.\ko \les \kosq(\epsc-\epsdt)^2 + 8\,(\piom)^2
\,(\epsc+\epsdt)\ris^{1/2}
\ric^{1/2}\, ;
\end{array}
\end{equation}
while
\begin{equation}
\epsdt = \frac{\epsa\epsb}{\epsa\cos^2\chi + \epsb\sin^2\chi}
\end{equation}
is defined for convenience.
We assume here that $\Omega$ is finite and exclude the possibility of
excitation of
axially propagating Voigt waves \c{L98}.

The summation symbols in \r{efield0} and \r{hfield0} are primed
to indicate that two of
the four modal coefficients $a_n$, ($1 \leq n \leq 4$), must be identically null-valued
in the present context. The determination of which
two requires analysis (and computation) of the $z$-directed components
of the modal time-averaged Poynting vectors $\#P_n(z)$, ($1 \leq n \leq 4$); thus,
\begin{eqnarray}
\nonumber
P_{nz}(z) &=&  \uz\. \#P_n(z)\\ [5pt]
&=&\frac{1}{2}\, \vert a_n\vert^2\, \exp\lec - 2\,{\rm Im}\les g_n\ris z\ric \,
{\rm Re} \les e_{n1}h_{n2}^\ast
- e_{n2}h_{n1}^\ast\ris\, ; \quad 1 \leq n \leq 4\, ,
\end{eqnarray}
where the asterisk denotes the complex conjugate. Because
$P_{1z} > 0$ and $P_{3z} < 0$, in general \c{L99},
we must have $a_1 \neq 0$ and $a_3 \equiv 0$. The
quantities $P_{2z}$ and $P_{4z}$ are always opposite
in sign; hence, either $a_2 \equiv 0$ when $P_{2z}/\vert a_2\vert^2 < 0$,
or $a_3 \equiv 0$ when $P_{4z}/\vert a_4\vert^2 < 0$. This
process also ensured,  for all calculations presented here,
that ${\rm Im}[g_1] \geq 0$ as well
as that either ${\rm Im}[g_2] \geq 0$  or ${\rm Im}[g_4] \geq 0$,
as appropriate.

The boundary value problem can now be formulated by ensuring the continuity
of the tangential components of the electric and the magnetic field phasors
across the plane $z=0$. The following four algebraic equations emerge:
\begin{eqnarray}
\label{equ1}
(\aal+\aar) + (\bbl+\bbr) &=& \sqrt{2}\,{\sum\limits_{n=1}^{4}}{'}\, a_n\,  e_{n1}\, ,\\
\label{equ2}
(\aal-\aar) - (\bbl-\bbr) &=& -\,\sqrt{2}i\,{\sum\limits_{n=1}^{4}}{'}\, a_n\,  e_{n2}\, ,\\
\label{equ3}
(\aal-\aar) + (\bbl-\bbr) &=& \sqrt{2}i\etao\,{\sum\limits_{n=1}^{4}}{'}\, a_n\,  h_{n1}\, ,\\
\label{equ4}
(\aal+\aar) - (\bbl+\bbr) &=& \sqrt{2}\etao\,{\sum\limits_{n=1}^{4}}{'}\, a_n\,  h_{n2}\, .
\end{eqnarray}
Their solution yields the four coefficients $\rrr$, etc.

\section{Numerical Results and Discussion}

All four reflection coefficients were
computed using Mathematica 3.0 on a Power
Macintosh 7300/180 computer. The values $h=1$,
$\lambdao = 600$~nm, $\chi = 30^\circ$ and $u = 10$ were fixed
for all calculations; while the volumetric fraction
$f$ and the structural half-period $\Omega$ were
varied. For a specified $\epss$, first $\epsa=\epsc$
and $\epsb$ were computed by solving \r{brug} iteratively;
and then the reflection coefficients $r_L$ and $r_R$
were obtained from the simultaneous
solution of \r{equ1}-\r{equ4}.
The relative permittivity scalars
$\epsa$ and $\epsb$ were plotted as functions of $f$, and the reflectances
\begin{equation}
R_{pq} = \Big\vert r_{pq}\Big\vert^2\, ; \quad p = L,\,R\,;\quad q = L,\,R\, ,
\end{equation}
as functions of both $f$ and $\Omega$.

In order to understand the response of a metallic TFHBM halfspace, it is
best to begin with the results for a virtually lossless dielectric
TFHBM hafspace.
Shown in Figure 1 are the computed values of $R_{RR}$, $R_{LL}$
and $R_{LR} = R_{RL}$~---~along with $\epsa$ and $\epsb$~---~when
$\epss=5+i0.001$. The TFHBM chosen is structurally right--handed;
and the CBP is manifested (i) as the very high-magnitude ridge in the
plot of $R_{RR}$ as well as (ii) the absence of that feature
in the plot of $R_{LL}$. This manifestation had been
observed earlier for axially excited, lossless dielectric TFHBM layers
\c{VL98a,VL98b}, but not for halfspaces. The CBP vanishes
when either $f=0$ or $f=1$, because the TFHBM is then 
isotropic (i.e., $\epsa=\epsb=\epsc$) and, therefore, homogeneous.

Next let us delineate the effect of absorption. The calculations for
Figure 1 were repeated, but with $\epss = 5+i0.5$. As the
volumetric fraction $f$ increases, absorption becomes
increasingly significant, as shown by the plots of $\epsa$ and
$\epsb$ in Figure 2. The CBP still leaves a broadly recognizable
signature~---~as also for absorbing TFHBM layers \c{L99m}~---~but
the initially high-magnitude ridge in the plot of $R_{RR}$ 
dissipates fairly rapidly
as $f$ increases from zero. Indeed, the CBP virtually disappears  
for $f \stackrel{\textstyle >}{\sim} 0.8$.

Finally, in Figure 3 are shown the calculated
results for a metallic TFHBM halfspace. The chosen value $\epss=-50+i18$
is very close to the relative permittivity of
bulk aluminum at $\lambdao=600$~nm \c{data}. The CBP has a
very clear signature in the plot of $R_{RR}$ 
in Figure 3, but only for very low values of $f$, when
the TFHBM is effectively an absorbing dielectric material. The high-magnitude
ridge broadens relative to the one in Figure 2, and vanishes for 
$f \stackrel{\textstyle >}{\sim} 0.4$. 
The reason is the
occurrence of percolation in the two-phase composite
material that the chosen TFHBM is \c{NSC}. 

Two absorbing-dielectric-to-metal transitions are evident in the
plots of $\epsa$ and $\epsb$ in Figure 3. The first transition
occurs at the low value $f \sim 0.05$, ${\rm Re}[\epsb]$ becoming
negative thereafter and ${\rm Im}[\epsb]$ increasing in magnitude
as $f$ increases therafter. The second transition occurs at $f \sim 0.5$,
with ${\rm Re}[\epsa] < 0$  thereafter and $\vert{\rm Im}[\epsa]\vert$ increasing in magnitude
as $f$ increases thereafter. 

The reason for the two transitions not
coinciding becomes clear on rewriting \r{epsbasic} and \r{epsref}
together as
\begin{equation}
\=\eps(\#r) = \epso\, \lec \epsa\,\les \=I -\#u_t(z)\#u_t(z) \ris
+ \epsb\,\#u_t(z)\#u_t(z) \ric\, ,
\end{equation}
wherein  the
equality $\epsc=\epsa$ has been assumed. Percolation
along the direction parallel to $\#u_t(z)$ occurs at  lower threshold values
of $f$ because of the connectedness of the metallic needles
in that direction; whereas the lower connectivity
of the metallic phase in all other directions delays the
onset of percolation in those directions. As $f$ increases,
percolation
occurs last in any direction perpendicular to $\#u_t(z)$.
In Figure 3, the inter-transition $f$-regime is most clearly evident
as $0.4 \stackrel{\textstyle <}{\sim} f \stackrel{\textstyle <}{\sim} 0.6$
in the plot of $R_{LR}$, which becomes
virtually constant at a high magnitude for $f \stackrel{\textstyle >}{\sim} 0.6$.

Although only two different  absorbing-dielectric-to-metal transitions
are evident in the plots of $\epsa$ and $\epsb$
in Figure 3 because of the assumption that $\epsc=\epsa$,
three different transitions will occur for a metallic TFHBM for
which $\epsc\neq\epsa$. As $f$ will increase from 0, the first transition
will be exhibited by $\epsb$. The second and the third transitions will be
located at higher values of $f$, in consequence of the columnar microstructure
being locally aciculate. The latter two transitions will concide for columns
with circular cross-sections, but not for noncircular columns. However,
the reflection of normally incident
plane waves by metallic TFHBM halfspaces and layers shall show
evidence of only two transitions, because only (i) $\epsc$ and (ii) the
combination $\epsdt$ of $\epsa$ and $\epsb$
enter the expressions \r{efield0}-\r{g2}. We also
conclude from the presented results that metallic TFHBMs with only
low metallic volumetric fractions appear desirable,
if the aim is to exploit the CBP for a certain application.

\newpage

%%%%%%%%%%%%% Figure 1 begins %%%%%%%%%%%%%%
\begin{figure}[!htb]
\centering
\includegraphics[width=126mm]{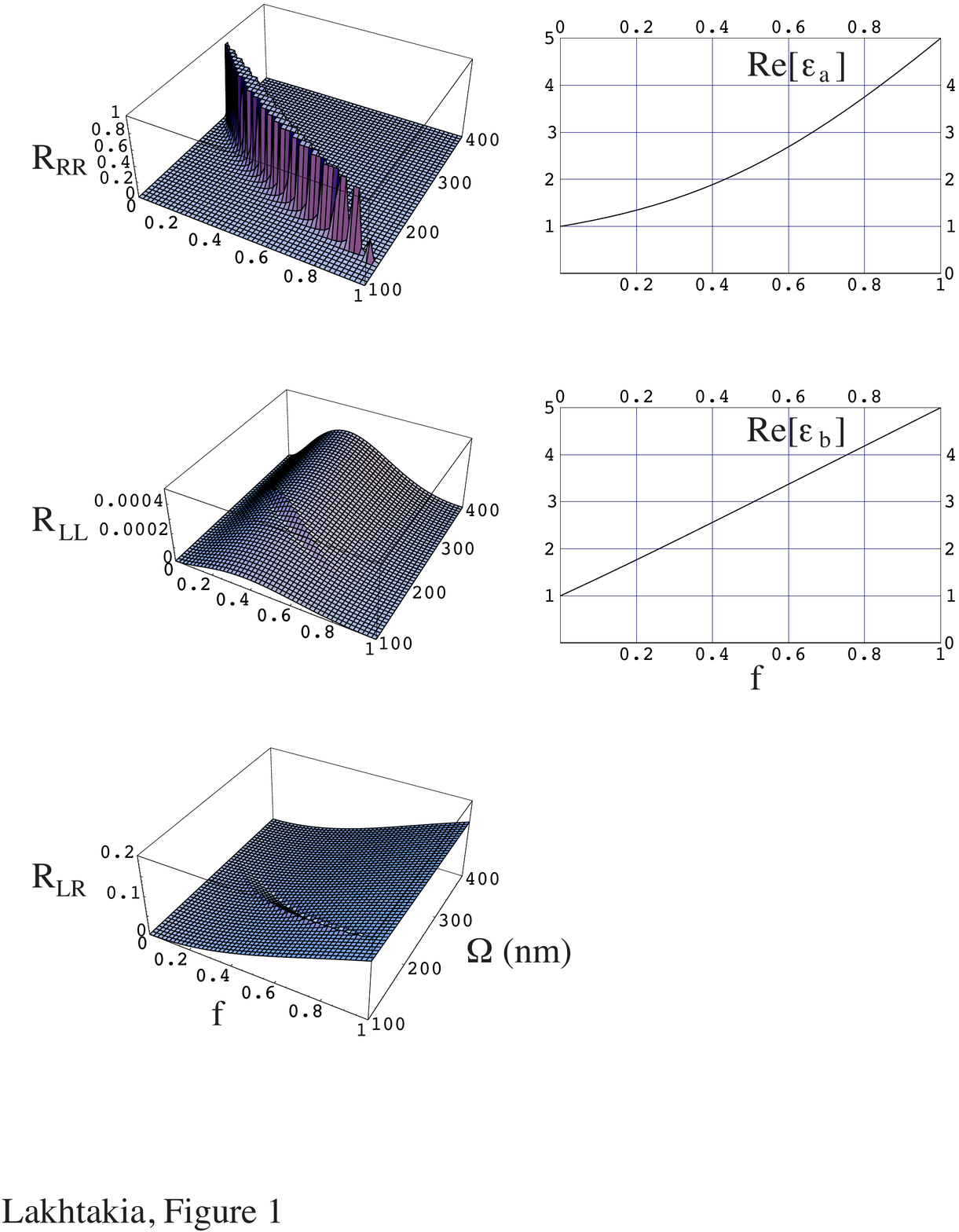}
\caption{Computed values of $R_{RR}$,
$R_{LL}$ and $R_{LR}=R_{RL}$ as functions
of $f$ and $\Omega$; as well as computed values of
$\epsa=\epsc$ and $\epsb$ as functions of $f$.
The TFHBM halfspace is made of a virtually lossless dielectric
material with $\epss = 5+i0.001$, so that ${\rm Im}[\epsa]$
and ${\rm Im}[\epsb]$ are too small to be shown. Other parameters are as
follows: $h=1$, $\lambdao=600$~nm, $\chi = 30^\circ$
and $u = 10$.
}\label{Fig1}
\end{figure}
%%%%%%%%%%%%% Figure 1 ends %%%%%%%%%%%%%%

%%%%%%%%%%%%% Figure 2 begins %%%%%%%%%%%%%%
\begin{figure}[!htb]
\centering
\includegraphics[width=126mm]{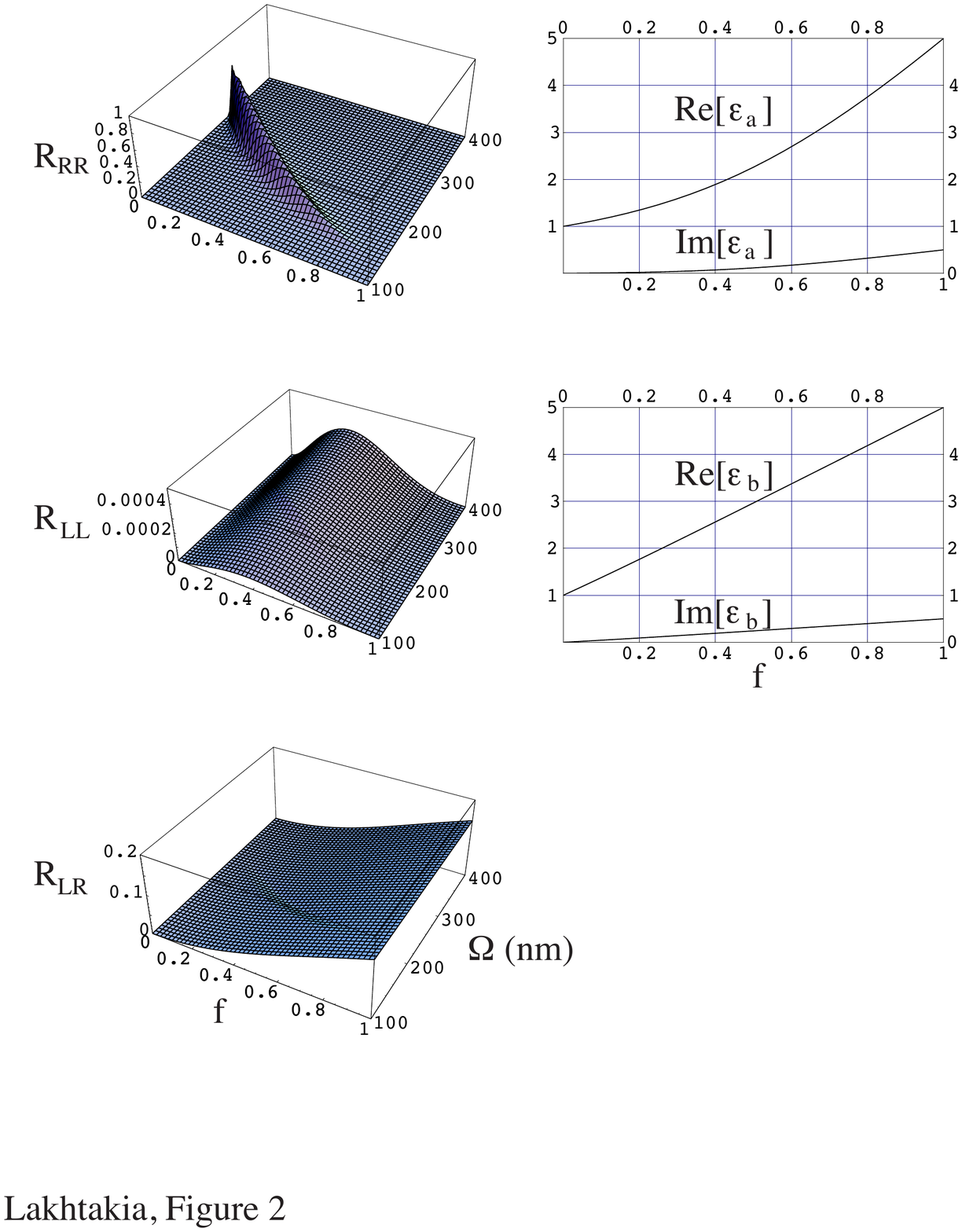}
\caption{Same as Figure 1, but
for a TFHBM halfspace made of an absorbing dielectric
material with $\epss = 5+i0.5$.
}\label{Fig2}
\end{figure}
%%%%%%%%%%%%% Figure 2 ends %%%%%%%%%%%%%%

%%%%%%%%%%%%% Figure 3 begins %%%%%%%%%%%%%%
\begin{figure}[!htb]
\centering
\includegraphics[width=126mm]{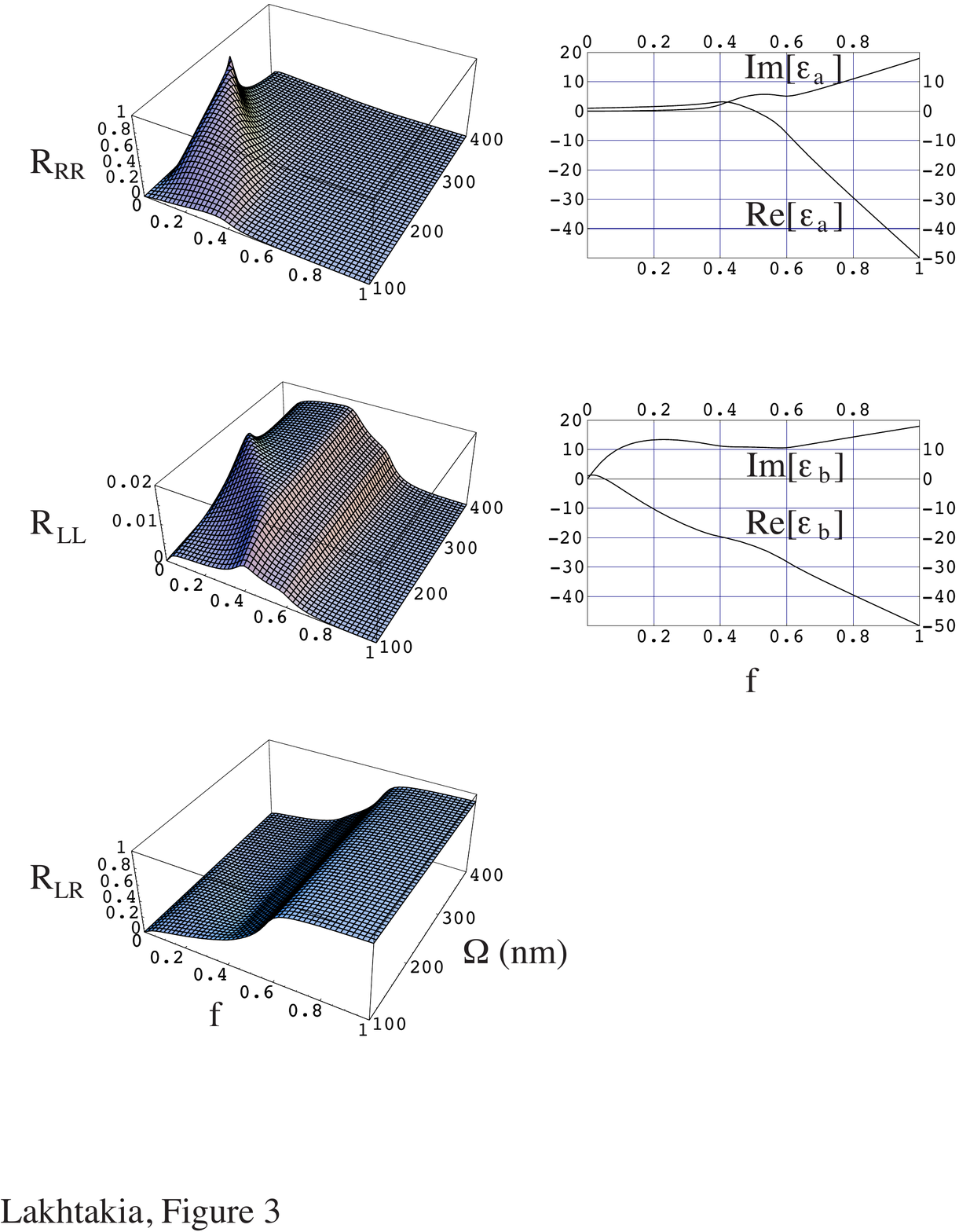}
\caption{Same as Figure 1, but
for a TFHBM halfspace made of a metal with $\epss = -50 + i18$. 
}\label{Fig3}
\end{figure}
%%%%%%%%%%%%% Figure 3 ends %%%%%%%%%%%%%%

\end{document}